\documentclass[prl,twocolumn,10pt,showpacs]{revtex4} 
\usepackage[centertags]{amsmath}
\usepackage{bbold} 
\usepackage{amsthm}
\usepackage[latin1]{inputenc} 
\newcommand{\tr}{\operatorname{tr}}  

\newcommand{\PROJ}[1]{\Zust{#1}\!\Bra{#1}}
\newcommand{\PR}[2]{\Zust{#1}\!\Bra{#2}}

\newcommand{\Zust}[1]{|#1\rangle}

\newcommand{\Bra}[1]{\langle#1|} 
\newcommand{\HR}{{\cal H}}

\newcommand{\CN}{{\mathbb C}}
\newcommand{\RN}{{\mathbb R}}

\newcommand{\diag}{{\text{diag}}} 

\newcommand{\EE}{{\cal E}}
\newcommand{\OO}{{\cal O}}
\renewcommand{\Re}{\operatorname{Re}}

\begin{document}
\title{A necessary and sufficient condition for optimal decompositions}
\author{Tobias Prager}
\affiliation{Institut für Theoretische Physik,TU Berlin, 10623 Berlin,
  Germany}
\email{tobias@physik.tu-berlin.de}
\begin{abstract}
 An important measure of bipartite entanglement is
  the entanglement of formation, which is defined as the
  minimum average pure state entanglement of all decompositions
  realizing a given  state. A decomposition which achieves this minimum is
  called an \emph{optimal} decomposition. However, as for the
  entanglement of formation, 
  there is not much known about the 
  structure of such optimal decompositions, except for some special
  cases, like states of two qubits or isotropic states.
  Here we present a  necessary and sufficient condition
  for a set of pure states of a finite dimensional bipartite systems
  to form an optimal decomposition. This
  condition is well suited  to treat the question, whether the
  entanglement of formation is additive or not.
\end{abstract}
\maketitle

One of the main ingredients in quantum information theory is
entanglement. Therefore, there have been many attempts to 
 understand this property better. While in the case of bi-partite pure
states, entanglement is well understood
and uniquely quantified under some
general assumptions \cite{BBPS} this is not the case for mixed
states. There have 
been many different entanglement measures proposed for such states,
one of which 
is the \emph{entanglement of formation} \cite{Bennet}, defined as
\begin{eqnarray}\label{eof}
  E_f(\rho):=\min\{\sum_i p_i E(\Zust{\psi_i}) \Big| \sum_i p_i
  \PROJ{\psi_i}=\rho\},
\end{eqnarray}
where $E$ denotes  the entanglement measure on
pure states,
\begin{eqnarray*}
  E(\Zust{\psi})=-\tr \sigma \log_2 \sigma, \quad
  \sigma=\tr_B \PROJ{\psi}.
\end{eqnarray*}
Closely related to the entanglement of formation  are so
called \emph{optimal decompositions} of a state, which are those
decompositions $\{(p_i,\Zust{\psi_i}\}_{i}$, that
achieve the minimum in eq. (\ref{eof}).
However, for all cases except $\CN^2\otimes\CN^2$ \cite{Wootters} 
and some highly
symmetric states in higher dimensions \cite{tv}, it is not
known, how to compute an optimal decomposition for a given state nor
the entanglement of formation itself. 

One property of the  entanglement of formation is being affine on the
convex set of states, generated by convex linear combinations of pure
states from an optimal
decomposition \cite{Uhl}. In other words, if the decomposition
$\{(p_i,\Zust{\psi_i}\}_{i=1}^n$ with $p_i>0$ 
 is optimal, then the decomposition 
$\{(q_i,\Zust{\psi_i}\}_{i=1}^n$ where the $q_i$ form an arbitrary
probability distribution is also optimal. The optimality of a
decomposition thus only depends on the states $\Zust{\psi_i}$  and not on the
corresponding probabilities. This is the justification to talk about the
optimality of a set of pure states $\{\Zust{\psi_i}\}_i$, by which we
  mean the optimality of the decompositions  $\{(p_i,\Zust{\psi_i})\}_i$.

A question concerning the entanglement of formation, which is widely
believed to be
true but not yet answered, is the question of additivity, i.e. whether
$E_f(\rho\otimes \sigma)=E_f(\rho)+E_f(\sigma)$ holds true for all
$\rho$ and $\sigma$. This question can be directly reformulated in terms of
decompositions: Given the optimal decompositions
$\{(p_i,\Zust{\psi_i})\}_{i=1}^n$ of $\rho$ and
$\{(q_j,\Zust{\phi_j})\}_{j=1}^m$ of $\sigma$, 
is the tensorproduct decomposition
$\{(p_i q_j ,\Zust{\psi_i}\otimes\Zust{\phi_j})\}_{i,j=1}^{n,m}$, which
is clearly a decomposition of $\rho \otimes \sigma$,
also
an \emph{optimal} one? 
In terms of optimal sets, this is equivalent to the question, whether the
set of pure states  $\{\Zust{\psi_i}\otimes\Zust{\phi_j}\}_{i,j=1}^{n,m}$
is optimal given that $\{\Zust{\psi_i}\}_{i=1}^n$ and
$\{\Zust{\phi_j}\}_{j=1}^m$ are optimal.
To address this question
we present a  necessary and sufficient condition,  to
decide whether a given set of pure states is optimal. 

Before we  continue, we need to fix some notations.
A decomposition of a state $\rho$ is normally defined as a
 set of normalized vectors $\Zust{\psi_i}$ 
with appropriate probabilities $p_i$ such that $\rho=\sum_{i} p_i
\PROJ{\psi_i}$.
However, for computational and notational purposes it is often
convenient 
to define a decomposition to be  a set  of non normalized vectors,
whose squared norms  correspond to the probabilities,
i.e. we make the transition
$\{(p_i,\Zust{\psi_i})\}_{i}\rightarrow \{\Zust{\tilde \phi_i}\}_{i}$
with $\Zust{\tilde \phi_i}=\sqrt{p_i}
  \Zust{\psi_i}$.
To distinguish non normalized from  normalized vectors we mark them
with a tilde. 
In this notation we have $\rho=\sum_i \PROJ{\tilde \phi_i}$ and
\begin{eqnarray*}
  E_f(\rho)=\min\big\{ \sum_i  \EE(\Zust{\tilde \phi_i}) \big| \sum_i
   \PROJ{\tilde \phi_i}=\rho\big\},
\end{eqnarray*}
where we have introduced the homogeneously extended pure state
entanglement
\begin{eqnarray*}
  \EE(\Zust{\tilde\phi}):=\|\tilde\phi\|^2
  E(\frac{\Zust{\tilde\phi}}{\|\tilde\phi\|})
  =-\tr \sigma \log_2 \frac{\sigma}{\tr\sigma}
\end{eqnarray*}
with $\sigma=\tr_B \PROJ{\tilde\phi}$.

To find  all possible decompositions $\{(q_j,\Zust{\phi_j})\}_{j=1}^m$
of a state $\rho$ from a given
decomposition $\{(p_i,\Zust{\psi_i})\}_{i=1}^n$
( e.g. the spectral decomposition) we will make repeated use of a
theorem by Wootters et al. \cite{JW}. This theorem  states, that two
decompositions $\{(q_j,\Zust{\phi_j})\}_{j=1}^m$
and $\{(p_i,\Zust{\psi_i})\}_{i=1}^n$
generate the same state if and only if they are related
by a right unitary matrix $U$, i.e. a matrix obeying
$U^\dagger U=\openone$,
in the following way:
\begin{eqnarray}
  \sqrt{q_j}\Zust{\phi_j}=\sum_{i=1}^n U_{ji} \sqrt{p_i}\Zust{\psi_i}
\end{eqnarray} 
or in terms of non normalized decompositions
\begin{eqnarray}\label{aes}
  \Zust{\tilde \phi_j}=\sum_{i=1}^n U_{ji} \Zust{\tilde \psi_i}.
\end{eqnarray}

We now state our main result, a necessary and sufficient 
 condition for the optimality of a set of pure states on a bipartite
 finite dimensional Hilbert space: 

\textbf{Theorem:}
Let $\Zust{\psi_i}\in \HR_A\otimes \HR_B$, $i=1,\ldots,n$  and
$\sigma_{ij}:=\tr_B \PR{\psi_i}{\psi_j}$. 
The set of pure states $\{\Zust{\psi_i}\}_{i=1}^n$ is optimal if and only if
  \begin{eqnarray}\label{newcondeq}
    \EE(\sum_{i=1}^n c_i \Zust{\psi_i})\ge -\Re\big(\sum_{i,j=1}^n c_i \bar c_j
      \tr[\sigma_{ij} \log_2 \sigma_{ii}]\big)
  \end{eqnarray}
holds true for arbitrary  $c_i \in \CN$.

\begin{proof}
First we prove
sufficiency: Let $\{\Zust{\psi_i}\}_{i=1}^n$ be a set of pure states for
which eq. (\ref{newcondeq}) is satisfied. 
Choosing an arbitrary probability distribution $\{p_i\}_{i=1}^n$ we
can form a decomposition $\{(p_i,\Zust{\psi_i})\}_{i=1}^n$ representing the
state $\rho=\sum_{i=1}^n p_i \PROJ{\psi_i}$.
Every other decomposition
$\{(q_j,\Zust{\phi_j})\}_{j=1}^m$ of $\rho$ 
is obtained via a right unitary matrix $U$
by 
\begin{eqnarray*}
  \sqrt{q_j} \Zust{\phi_j}=\sum_{i=1}^n U_{ji} \sqrt{p_i} \Zust{\psi_i}.
\end{eqnarray*}
From eq. (\ref{newcondeq}) we get
\begin{eqnarray*}
  \EE(\sqrt{q_j}\Zust{\phi_j})&=&
  \EE(\sum_{i=1}^n U_{ji} \sqrt{p_i} \Zust{\psi_i})\\
  &\ge&-\Re \Big(\sum_{i,k=1}^n U_{ji} \bar U_{jk}
      \sqrt{p_i p_k} \tr [\sigma_{ik} \log_2 \sigma_{ii}]\Big)
\end{eqnarray*}
with $\sigma_{ik}=\tr\PR{\psi_i}{\psi_k}$. Summing over $j$ and using
the right unitarity of $U$ we arrive at
\begin{eqnarray*}
  \lefteqn{\sum_{j=1}^m q_j E(\Zust{\phi_j})=\sum_{j=1}^m
  \EE(\sqrt{q_j}\Zust{\phi_j})}\\
  &\ge& -\Re \sum_{i=1}^n p_i \tr [\sigma_{ii} \log_2 \sigma_{ii}]=
  \sum_{i=1}^n p_i E(\Zust{\psi_i}) 
\end{eqnarray*}
which proves the optimality of the decomposition 
$\{(p_i,\Zust{\psi_i})\}_{i=1}^n$ and therefore the optimality of the
set $\{\Zust{\psi_i}\}_{i=1}^n$.

Next we prove necessity: Let $\{\Zust{\psi_i}\}_{i=1}^n$ be an
optimal set of pure states. 
Then  the  decomposition
$\{(\frac{1}{n},\Zust{\psi_i})\}_{i=1}^{n}
=:\{\Zust{\tilde \psi_i}\}_{i=1}^{n}
$ representing the state $\rho=\sum_{i=1}^n \frac{1}{n} \PROJ{\psi_i}$
is optimal. We now define a family of
unitary $(n+1)\times (n+1)$-matrizes by $\tilde U(t):=\exp(tT)$, where t is a
real parameter and $T$ is
the skew hermitian matrix defined by $T_{n+1,i}=c_i$, $T_{i,n+1}=-\bar
c_i$ and $T_{i,j}=0$ otherwise, i.e.

\begin{eqnarray*}
  T=\left(
    \begin{array}{ccccc}
      0 & 0 &\cdots & 0 &-\bar c_1\\
      0 & 0 &\cdots & 0 &-\bar c_2\\
      \vdots &\vdots& \ddots & \vdots &\vdots\\
      0 & 0 &\cdots & 0 &-\bar c_n\\
      c_1 & c_2& \cdots & c_n& 0
    \end{array} 
 \right).
\end{eqnarray*}
Selecting the first $n$  columns 
of $\tilde U(t)$ we get
a right unitary $(n+1)\times n$-matrix $U(t)$.
Applying $U(t)$ to the optimal decomposition
$\{\Zust{\tilde \psi_i}\}_{i=1}^{n}$  according to eq. (\ref{aes})
we get a new decomposition
$\{\Zust{\tilde \phi_j(t)}\}_{j=1}^{n+1}$ of $\rho$ which is defined  by
\begin{eqnarray}
  \Zust{\tilde \phi_i(t)}&=&\Zust{\tilde \psi_i}- \frac{t^2}{2}
  \sum_{j=1}^{n} c_j \bar c_i \Zust{\tilde \psi_j} +\OO(t^3),\;
  i=1,\ldots n \nonumber\\
  \Zust{\tilde \phi_{n+1}(t)}&=&t \sum_{i=1}^n  c_i 
  \Zust{\tilde \psi_i}+\OO(t^3)\label{phii}.    
\end{eqnarray}
This decomposition can  not  have less entanglement than the
decomposition 
$\{\Zust{\tilde \psi_i}\}_{i=1}^{n}$, which was assumed to be optimal,
thus
\begin{eqnarray}\label{oes}
  \sum_{i=1}^{n} \EE(\Zust{\tilde\psi_i})\le
  \sum_{j=1}^{n+1} \EE\big(\Zust{\tilde\phi_j(t)}\big) \quad \forall t\in \RN
\end{eqnarray}
holds true. At $t$=0, both sides of this inequality are equal. In the
following, we will show that the first derivatives with respect to $t$
of both sides are also equal at $t=0$. Hence, the inequality must be fullfilled
for the second derivatives at $t=0$ which will lead to eq (\ref{newcondeq}).
To proceed, we need the following lemma:

\textbf{Lemma: } 
  Let $A(t)$ be a family of positive matrices, 
which depend differentiably on $t\in(a,b)$. Then we have
  \begin{eqnarray}\label{fdhe}
    -\frac{d}{dt}\tr [A(t) \log_2 \frac{A(t)}{\tr A(t)}]=
 -\tr [A(t)' \log_2 \frac{A(t)}{\tr A(t)}]
  \end{eqnarray}
Additionaly, if $\frac{d}{dt} A(t)$ is zero  at some point $t_0$  and
if $A(t)$ is twice differentiable at $t_0$
we have
\begin{equation}\label{sdhe}
  \begin{split}
  -\frac{d^2}{dt^2}\tr [A(t) \log_2 \frac{A(t)}{\tr A(t)}]\Big|_{t=t_0}\!\!\!\!=
  -\tr [A(t_0){''} \log_2 \frac{A(t_0)}{\tr A(t_0)}].    
  \end{split}
\end{equation}
Note that eq. (\ref{fdhe}) is always well defined due to the positivity
and differentiability of $A(t)$ whereas eq. (\ref{sdhe}) may not be
well defined.

\begin{proof}
Let $A(t)=U(t) \diag(\vec{\lambda}(t)) U^\dagger(t)$ with $U(t)$ unitary.
Then we have
\begin{eqnarray*}
  \lefteqn{-\frac{d}{dt}\tr [A(t)\log_2 \frac{A(t)}{\tr A(t)}]}\\
  &=&-\frac{d}{dt}\big(\sum_i \lambda_i(t)(
  \log_2 \lambda_i(t) -\log_2 \sum_j \lambda_j(t))\big)\\
  &=&-\sum_i \lambda_i'(t)(
  \log_2 \lambda_i(t) -\log_2 \sum_j \lambda_j(t))\\
  &=&-\tr [A'(t)(
  \log_2 A(t) -\log_2 \tr A(t))]\\
  &=&-\tr [A(t)' \log_2 \frac{A(t)}{\tr A(t)}].
\end{eqnarray*}
In the last but one step,
we have used that 
$U'(t)U^\dagger(t)+U(t) U^{\dagger \prime}(t)=0$, that
the trace is invariant under cyclic permutations of the factors and
that $\diag(\vec\lambda(t))$ and $\log_2[\diag(\vec\lambda(t))]$
commute.
To prove the second part of the lemma, we use the fact 
 that if $A'(t_0)$ is zero $\lambda_i'(t_0)$ is zero and we
can always choose $U(t)$  such that $U'(t_0)=0$. This choice leads to
\begin{eqnarray*}
  \lefteqn{-\frac{d^2}{dt^2}\tr [A(t) \log_2 \frac{A(t)}{\tr
      A(t)}]
    \Big|_{t=t_0}=}\\
  &=&-\sum_i \lambda_i{''}(t_0)(
  \log_2 \lambda_i(t_0) -\log_2 \sum_j \lambda_j(t_0))\\
  &=&-\tr [A{''}(t_0)
  \log_2 \frac{A(t_0)}{\tr A(t_0)}].
\end{eqnarray*}
having used
$U''(t_0)U^\dagger(t_0)+U(t_0) U^{\dagger\prime\prime}(t_0)=0$.
which holds true, because $U'(t_0)=0$.
\end{proof}

Applying this lemma to $\tilde \pi_i(t):=\tr_B \PROJ{\tilde
  \phi_i(t)}$, $i=1,\ldots,n$,
we get
\begin{eqnarray*}
\frac{d}{d t}
  \EE(\Zust{\tilde \phi_i(t)})&=&
-\frac{d}{dt}\big( \tr \tilde \pi_i(t) \log_2 \frac{\tilde \pi_i(t)}{\tr \tilde
   \pi_i(t)}\big)\\
&=&-\tr \tilde \pi'_i(t) \log_2 \frac{\tilde \pi_i(t)}{\tr \tilde
   \pi_i(t)}.
\end{eqnarray*}
According to eq. (\ref{phii}), $\tilde \pi_i'(t)$ is zero at $t=0$. 
Therefore we can calculate the second derivative
of $\EE(\Zust{\tilde \phi_i(t)})$ at $t=0$ using the second part of the lemma, 
which leads to
\begin{eqnarray*}
\frac{d^2}{d t^2}
  \EE(\Zust{\tilde \phi_i(t)})\Big|_{t=0}&=&-\tr \tilde \pi{''}_i(0) 
  \log_2 \frac{\tilde \pi_i(0)}{\tr \tilde
   \pi_i(0)}
\end{eqnarray*}
Using  eq. (\ref{phii}) we can further evaluate the righthand side.
Defining $\tilde\sigma_{ij}:=\tr_B \PR{\tilde \psi_i}{\tilde \psi_j}$,
we arrive at
\begin{eqnarray}\label{sd1}
 \frac{d^2}{d t^2}
  \EE(\Zust{\tilde \phi_i(t)})\big|_{t=0}
&=&\sum_{j=1}^n \tr
  [(c_i \bar c_j \tilde\sigma_{ij} +c_j \bar c_i \tilde\sigma_{ji})\log_2
  \frac{\tilde\sigma_{ii}}{\tr \tilde\sigma_{ii}}]\nonumber\\
&=&2\Re\Big(\sum_{j=1}^n c_i \bar c_j \tr[\tilde\sigma_{ij}\log_2
  \frac{\tilde\sigma_{ii}}{\tr \tilde\sigma_{ii}}]   \Big)\nonumber\\
&=&\frac{2}{n}\Re\Big(\sum_{j=1}^n c_i \bar c_j \tr[\sigma_{ij}\log_2
  \sigma_{ii}]   \Big).
\end{eqnarray}
This expression is always well defined  because
$\tr \sigma_{ij} \log_2 \sigma_{ii}$ is well defined, even if
$\sigma_{ii}$ has a non vanishing kernel. This can  be
easily seen by expressing $\Zust{\psi_i}$ and $\Zust{\psi_j}$ in the
Schmidt-basis of $\Zust{\psi_i}$ and then evaluating 
$\tr \sigma_{ij} \log_2 \sigma_{ii}$ in this basis.

To calculate the derivatives of
$\EE(\Zust{\tilde \phi_{n+1}(t)})$ we use the fact, that $\EE(c
\Zust{\tilde \eta})=|c|^2 \EE(\Zust{\tilde\eta})$ which leads to 
\begin{eqnarray*}    
  \frac{d}{dt}
  \EE(\Zust{\tilde\phi_{n+1}(t)})\big|_{t=0}&=&\frac{d}{dt}[ t^2
  \EE(\sum_{i=1}^n c_i \Zust{\tilde\psi_i}+\OO(t^2))]\big|_{t=0}\nonumber\\&=&0
\end{eqnarray*}
and
\begin{eqnarray}\label{sd2}    
  \frac{d^2}{d t^2}
  \EE(\Zust{\tilde\phi_{n+1}(t)})\big|_{t=0}&=&\frac{d^2}{d t^2}[ t^2
  \EE(\sum_{i=1}^n c_i \Zust{\tilde\psi_i}+\OO(t^2))]\big|_{t=0}\nonumber \\
  &=&\frac{2}{n}
  \EE(\sum_{i=1}^n c_i \Zust{\psi_i}).
\end{eqnarray}
Note that the  first and second derivative of 
$\EE(\sum_{i=1}^n c_i \Zust{\tilde\psi_i}+\OO(t^2))]\big|_{t=0}$ are
well defined.

As we have shown, the first derivatives of the right and left hand side of
eq. (\ref{oes}) are equal to 0.
Therefore 
\begin{eqnarray}\label{sd}
  \frac{d^2}{d t^2}\Big(
  \sum_{i=1}^{n} \EE(\Zust{\tilde\psi_i})\Big)\Big|_{t=0}\le
  \frac{d^2}{d t^2}\Big(
  \sum_{j=1}^{n+1} \EE\big(\Zust{\tilde\phi_j(t)}\big)\Big)\Big|_{t=0}
\end{eqnarray}
necessarily holds true,
which in turn can be seen to be equivalent to
eq. (\ref{newcondeq}) by substituting eqs. (\ref{sd1}) and (\ref{sd2})
into the right hand side 
as well as taking into account, that the left hand side of eq. (\ref{sd}) 
does not depend on $t$ and
therefore is zero.
\end{proof}

Our condition for the optimality of a set of pure states is
a generalization of a result by Benatti and Narnhofer \cite{BN},
who proved a similar condition for the special case
of a decompositions consisting  only of
\emph{two} vectors.

It is easily seen, that it is sufficient to demand
eq. (\ref{newcondeq}) to be true for all $c_i\in \CN$ with $\|\vec c\|=1$
because we can cancel the square of the 
norm of the coefficient vector $\vec c$ on
both sides of the inequality. 
Additionally, \emph{if} the set $\{\Zust{\psi_i}\}_{i=1}^n$ is
optimal, the sum on the right side of eq. (\ref{newcondeq}) is already
real. This is implied by the fact, that for two states $\Zust{\psi_k}$
and $\Zust{\psi_l}$ from an optimal set, we have
\begin{eqnarray} \label{nc}
  \tr \sigma_{kl} \log_2 \sigma_ {kk}=
  \overline{\tr \sigma_{lk} \log_2 \sigma_{ll}},\;
  \sigma_{kl}:=\tr_B\PR{\psi_k}{\psi_l}. 
\end{eqnarray}
To prove this equation  we  define the decompositions 
$\{\Zust{\tilde \phi_i(\theta)}\}_{i=1}^n$
with 
\begin{eqnarray*}
\Zust{\tilde\phi^\pm_k(\theta)}&=&\frac{1}{\sqrt{n}}(
\cos \theta \Zust{\psi_k}\pm\sin \theta
\Zust{\psi_l})\\
\Zust{\tilde\phi^\pm_l(\theta)}&=&\frac{1}{\sqrt{n}}(\mp\sin\theta
\Zust{\psi_k}+
\cos \theta
\Zust{\psi_l})\text{  and }\\
\Zust{\tilde\phi^\pm_i(\theta)}&=&\frac{1}{\sqrt{n}} \Zust{\psi_i}
\quad i\neq k,l  
\end{eqnarray*}
 and $\{\Zust{\tilde \eta_i(\theta)}\}_{i=1}^n$
with 
\begin{eqnarray*}
\Zust{\tilde\eta^\pm_k(\theta)}&=&\frac{1}{\sqrt{n}}(
\cos \theta \Zust{\psi_k}\pm I\sin \theta
\Zust{\psi_l})\\
\Zust{\tilde\eta^\pm_l(\theta)}&=&\frac{1}{\sqrt{n}}(\pm I\sin\theta
\Zust{\psi_k}+
\cos \theta
\Zust{\psi_l})\text{  and }\\
\Zust{\tilde\eta^\pm_i(\theta)}&=&\frac{1}{\sqrt{n}} \Zust{\psi_i}
\quad i\neq k,l.  
\end{eqnarray*}
For $\theta=0$ those decompositions are identical to 
$\{\frac{1}{\sqrt{n}}\Zust{\psi_i}\}_{i=1}^n$ which is 
optimal by assumption.
Hence
\begin{eqnarray*}
  \frac{d}{d\theta} \sum_{i=1}^n
  \EE(\Zust{\tilde\phi^\pm_i(\theta)})\Big|_{t=0}\ge 0\text{ and } 
  \frac{d}{d\theta} \sum_{i=1}^n \EE(\Zust{\tilde\eta^\pm_i(\theta)})\Big|_{t=0}\ge0
\end{eqnarray*}
have to hold true, implying eq. (\ref{nc}) by using the lemma 
and some basic algebra.

This theorem is usefull to verify the optimality of a given
decomposition or a given set of pure states, because 
we only have to optimize a function over the unit sphere in $\CN^n$.
In contrast, by using  the straight forward condition
\begin{eqnarray}\label{oldcond}
  \sum_{i=1}^n \EE(\Zust{\tilde \psi_i})\le \sum_{j=1}^m \EE(\sum_{i=1}^n
  u_{ji}\Zust{\tilde\psi_i}) \; \forall \; U \text{ with } U^\dagger U=\mathbf{1}
\end{eqnarray}
for the optimality of
the decomposition $\{\Zust{\tilde\psi_i}\}_{i=1}^n$ which is easily derived
from the definition of an optimal decomposition and eq. (\ref{aes}),
we had to optimize a function over the
set of all right unitary $m\times n$-matrices. In this case  
$m$ must be at least $[\operatorname{rank}\rho]^2$ with 
$\rho=\sum_{i=1}^n \PROJ{\tilde \psi_i}$, which is the 
lowest known bound on the number
of pure states in an optimal decomposition \cite{Uhlmann}. 
Clearly this task is harder to perform.
However, as there is nothing for
free, the information we get in both cases is different: In doing the
optimization over the right unitary matrizes we arrive at an optimal
decomposition from which the entanglement of formation can be easily
calculated. In doing the optimization over the unit sphere in $\CN^n$
, we only get
the information whether the decomposition was optimal or not. 
Therefore our condition is of no direct use in finding  optimal
decompositions and thus calculating the entanglement of formation, but
there is another important application,
namely testing  the additivity of the entanglement of formation. 
To perform this task, we have to check, whether the set of 
tensorproduct states 
$\{\Zust{\psi^1_i}\otimes \Zust{\psi^2_j}\}_{i,j=1}^{n,m}$ is optimal,
given two optimal sets of states $\{\Zust{\psi^1_i}\}_{i=1}^n$ and 
$\{\Zust{\psi^2_j}\}_{j=1}^{m}$. Numerically this task is
easier to perform  using our theorem, than using 
eq.(\ref{oldcond}).
It might even be possible,
that the new  condition is helpful in  proving additivity of the
entanglement of 
formation analytically.

Another information we get from our condition is an estimation of the
entanglement of superpositions of states from an optimal set 
in terms of the
entanglement of the  states themselves and some overlap-terms.
This can be seen by
writing  the diagonal terms in eq. (\ref{newcondeq}) separately, 
  \begin{eqnarray*}
   \lefteqn{ \EE(\sum_{i=1}^n c_i \Zust{\psi_i})\ge \sum_{i=1}^n |c_i|^2
  E(\Zust{\psi_i})-}\\
 &&\qquad-\Re\big(\sum_{\substack{i,j=1\\i\neq j}}^n c_i \bar c_j
      \tr[\sigma_{ij} \log_2 \sigma_{ii}]\big)
    \quad \forall \; c_i \in \CN.
  \end{eqnarray*} 

Finally we conclude that in some sense locally optimal
decompositions are already optimal:
In deriving eq. (\ref{newcondeq}) we have only  
used derivatives and the value itself 
at the point $t=0$ in eq. (\ref{oes}). 
Therefore it is  sufficient to demand eq.
(\ref{oes}) to be true only in a small neighbourhood of $t=0$. This in
turn is
equivalent to demanding, that the decompositions 
$\{\Zust{\tilde \psi_i}\}_{i=1}^{n}$
has lower average entanglement than every decomposition, 
which we get from
it by a right unitary $(n+1)\times n$-matrix, which consists of the first $n$
columns of a unitary $(n+1)\times (n+1)$-matrix
in an arbitrary small neighbourhood of the identity, 
i.e. demanding local optimality. But since eq. (\ref{newcondeq}) is
also a sufficient condition for optimality, we conclude that local
optimality implies optimality.

The author would like to thank A. Szkola, C. Witte and
K.E. Hellwig  for
usefull discussions. The author gratefully acknowledges 
financial support from the
Deutsche Forschungsgemeinschaft (DFG).

\end{document}